
\magnification=\magstep1

\newbox\SlashedBox
\def\slashed#1{\setbox\SlashedBox=\hbox{#1}
\hbox to 0pt{\hbox to 1\wd\SlashedBox{\hfil/\hfil}\hss}#1}
\def\hboxtosizeof#1#2{\setbox\SlashedBox=\hbox{#1}
\hbox to 1\wd\SlashedBox{#2}}

\def\mathslashed#1{\setbox\SlashedBox=\hbox{$#1$}
\hbox to 0pt{\hbox to 1\wd\SlashedBox{\hfil/\hfil}\hss}#1}

\def\ifsmall{\iffalse}  
\def\titlepagefont{}  

\def\DefineTeXgraphics{%
\special{ps::[global] /TeXgraphics { } def}}  

\def\today{\ifcase\month\or January\or February\or March\or April\or May
\or June\or July\or August\or September\or October\or November\or
December\fi\space\number\day, \number\year}
\def\eatPrefix19{}
\def\Year{\expandafter\eatPrefix\the\year}
\newcount\hours \newcount\minutes
\def\monthname{\ifcase\month\or
January\or February\or March\or April\or May\or June\or July\or
August\or September\or October\or November\or December\fi}
\def\shortmonthname{\ifcase\month\or
Jan\or Feb\or Mar\or Apr\or May\or Jun\or Jul\or
Aug\or Sep\or Oct\or Nov\or Dec\fi}

\def\TimeStamp{\hours\the\time\divide\hours by60%
\minutes -\the\time\divide\minutes by60\multiply\minutes by60%
\advance\minutes by\the\time%
${\rm \shortmonthname}\cdot\if\day<10{}0\fi\the\day\cdot\the\year%
\qquad\the\hours:\if\minutes<10{}0\fi\the\minutes$}







\newif\ifdraftmode
\newif\ifleftlabels  

\def\nolabels{\def\wrlabeL##1{}\def\eqlabeL##1{}\def\reflabeL##1{}}
\def\writelabels{\def\wrlabeL##1{\leavevmode\vadjust{\rlap{\smash%
{\line{{\escapechar=` \hfill\rlap{\sevenrm\hskip.03in\string##1}}}}}}}%
\def\eqlabeL##1{{\escapechar-1\rlap{\sevenrm\hskip.05in\string##1}}}%
\def\reflabeL##1{\noexpand\rlap{\noexpand\sevenrm[\string##1]}}}
\def\writeleftlabels{\def\wrlabeL##1{\leavevmode\vadjust{\rlap{\smash%
{\line{{\escapechar=` \hfill\rlap{\sevenrm\hskip.03in\string##1}}}}}}}%
\def\eqlabeL##1{{\escapechar-1%
\rlap{\sixrm\hskip.05in\string##1}%
\llap{\sevenrm\string##1\hskip.03in\hbox to \hsize{}}}}%
\def\reflabeL##1{\noexpand\rlap{\noexpand\sevenrm[\string##1]}}}
\nolabels

\newdimen\fullhsize
\newdimen\hstitle
\hstitle=\hsize 
\newdimen\hsbody
\hsbody=\hsize 
\newdimen\hbodyoffset
\hbodyoffset=\hoffset 
\newbox\leftpage
\def\abstract#1{#1}
\def\rotated{\special{ps: landscape}
\magnification=1000  
\baselineskip=14pt
\global\hstitle=9truein\global\hsbody=4.75truein
\global\vsize=7truein\global\voffset=-.31truein
\global\hoffset=-0.54in\global\hbodyoffset=-.54truein
\global\fullhsize=10truein
\def\DefineTeXgraphics{%
\special{ps::[global]
/TeXgraphics {currentpoint translate 0.7 0.7 scale
              -80 0.72 mul -1000 0.72 mul translate} def}}
\let\lr=L
\def\ifsmall{\iftrue}
\def\titlepagefont{\twelvepoint}
\trueseventeenpoint
\def\almostshipout##1{\if L\lr \count1=1
      \global\setbox\leftpage=##1 \global\let\lr=R
   \else \count1=2
      \shipout\vbox{\hbox to\fullhsize{\box\leftpage\hfil##1}}
      \global\let\lr=L\fi}

\output={\ifnum\count0=1 
 \shipout\vbox{\hbox to \fullhsize{\hfill\pagebody\hfill}}\advancepageno
 \else
 \almostshipout{\leftline{\vbox{\pagebody\makefootline}}}\advancepageno
 \fi}

\def\abstract##1{{\leftskip=1.5in\rightskip=1.5in ##1\par}} }

\def\linemessage#1{\immediate\write16{#1}}

\global\newcount\secno \global\secno=0
\global\newcount\appno \global\appno=0
\global\newcount\meqno \global\meqno=1
\global\newcount\subsecno \global\subsecno=0
\global\newcount\figno \global\figno=0

\newif\ifAnyCounterChanged
\let\terminator=\relax
\def\normalize#1{\ifx#1\terminator\let\next=\relax\else%
\if#1i\aftergroup i\else\if#1v\aftergroup v\else\if#1x\aftergroup x%
\else\if#1l\aftergroup l\else\if#1c\aftergroup c\else%
\if#1m\aftergroup m\else%
\if#1I\aftergroup I\else\if#1V\aftergroup V\else\if#1X\aftergroup X%
\else\if#1L\aftergroup L\else\if#1C\aftergroup C\else%
\if#1M\aftergroup M\else\aftergroup#1\fi\fi\fi\fi\fi\fi\fi\fi\fi\fi\fi\fi%
\let\next=\normalize\fi%
\next}
\def\makeNormal#1#2{\def\doNormalDef{\edef#1}\begingroup%
\aftergroup\doNormalDef\aftergroup{\normalize#2\terminator\aftergroup}%
\endgroup}

\def\warnIfChanged#1#2{%
\ifundef#1
\else\begingroup%
\edef\oldDefinitionOfCounter{#1}\edef\newDefinitionOfCounter{#2}%
\ifx\oldDefinitionOfCounter\newDefinitionOfCounter%
\else%
\linemessage{Warning: definition of \noexpand#1 has changed.}%
\global\AnyCounterChangedtrue\fi\endgroup\fi}

\def\Section#1{\global\advance\secno by1\relax\global\meqno=1%
\global\subsecno=0%
\bigbreak\bigskip
\centerline{\twelvepoint \bf %
\the\secno. #1}%
\par\nobreak\medskip\nobreak}
\def\tagsection#1{%
\warnIfChanged#1{\the\secno}%
\xdef#1{\the\secno}%
\ifWritingAuxFile\immediate\write\auxfile{\noexpand\xdef\noexpand#1{#1}}\fi%
}
\def\section{\Section}
\def\Subsection#1{\global\advance\subsecno by1\relax\medskip %
\leftline{\bf\the\secno.\the\subsecno\ #1}%
\par\nobreak\smallskip\nobreak}
\def\tagsubsection#1{%
\warnIfChanged#1{\the\secno.\the\subsecno}%
\xdef#1{\the\secno.\the\subsecno}%
\ifWritingAuxFile\immediate\write\auxfile{\noexpand\xdef\noexpand#1{#1}}\fi%
}

\def\subsection{\Subsection}

\def\romappno{\uppercase\expandafter{\romannumeral\appno}}
\def\makeNormalizedRomappno{%
\expandafter\makeNormal\expandafter\normalizedromappno%
\expandafter{\romannumeral\appno}%
\edef\normalizedromappno{\uppercase{\normalizedromappno}}}
\def\Appendix#1{\global\advance\appno by1\relax\global\meqno=1\global\secno=0
\bigbreak\bigskip
\centerline{\twelvepoint \bf Appendix %
\romappno. #1}%
\par\nobreak\medskip\nobreak}
\def\tagappendix#1{\makeNormalizedRomappno%
\warnIfChanged#1{\normalizedromappno}%
\xdef#1{\normalizedromappno}%
\ifWritingAuxFile\immediate\write\auxfile{\noexpand\xdef\noexpand#1{#1}}\fi%
}
\def\appendix{\Appendix}

\def\eqn#1{\makeNormalizedRomappno%
\ifnum\secno>0%
  \warnIfChanged#1{\the\secno.\the\meqno}%
  \eqno(\the\secno.\the\meqno)\xdef#1{\the\secno.\the\meqno}%
     \global\advance\meqno by1
\else\ifnum\appno>0%
  \warnIfChanged#1{\normalizedromappno.\the\meqno}%
  \eqno({\rm\romappno}.\the\meqno)%
      \xdef#1{\normalizedromappno.\the\meqno}%
     \global\advance\meqno by1
\else%
  \warnIfChanged#1{\the\meqno}%
  \eqno(\the\meqno)\xdef#1{\the\meqno}%
     \global\advance\meqno by1
\fi\fi%
\eqlabeL#1%
\ifWritingAuxFile\immediate\write\auxfile{\noexpand\xdef\noexpand#1{#1}}\fi%
}
\def\defeqn#1{\makeNormalizedRomappno%
\ifnum\secno>0%
  \warnIfChanged#1{\the\secno.\the\meqno}%
  \xdef#1{\the\secno.\the\meqno}%
     \global\advance\meqno by1
\else\ifnum\appno>0%
  \warnIfChanged#1{\normalizedromappno.\the\meqno}%
  \xdef#1{\normalizedromappno.\the\meqno}%
     \global\advance\meqno by1
\else%
  \warnIfChanged#1{\the\meqno}%
  \xdef#1{\the\meqno}%
     \global\advance\meqno by1
\fi\fi%
\eqlabeL#1%
\ifWritingAuxFile\immediate\write\auxfile{\noexpand\xdef\noexpand#1{#1}}\fi%
}
\def\anoneqn{\makeNormalizedRomappno%
\ifnum\secno>0
  \eqno(\the\secno.\the\meqno)%
     \global\advance\meqno by1
\else\ifnum\appno>0
  \eqno({\rm\normalizedromappno}.\the\meqno)%
     \global\advance\meqno by1
\else
  \eqno(\the\meqno)%
     \global\advance\meqno by1
\fi\fi%
}
\def\mfig#1#2{\global\advance\figno by1%
\relax#1\the\figno%
\warnIfChanged#2{\the\figno}%
\edef#2{\the\figno}%
\reflabeL#2%
\ifWritingAuxFile\immediate\write\auxfile{\noexpand\xdef\noexpand#2{#2}}\fi%
}

\def\fig#1{\mfig{fig.~}#1}

\catcode`@=11 

\font\ninerm=cmr9
\font\eightrm=cmr8
\font\sixrm=cmr6

\def\loadtrueseventeenpoint{
 \font\seventeenrm=cmr10 at 17.28truept
 \font\seventeeni=cmmi10 at 17.28truept
 \font\seventeenbf=cmbx10 at 17.28truept
 \font\seventeenit=cmti10 at 17.28truept
 \font\seventeensl=cmsl10 at 17.28truept
 \font\seventeensy=cmsy10 at 17.28truept
}
\def\loadfourteenpoint{
\font\fourteenrm=cmr10 at 14.4pt
\font\fourteeni=cmmi10 at 14.4pt
\font\fourteenit=cmti10 at 14.4pt
\font\fourteensl=cmsl10 at 14.4pt
\font\fourteensy=cmsy10 at 14.4pt
\font\fourteenbf=cmbx10 at 14.4pt
}
\def\loadtruetwelvepoint{
\font\twelverm=cmr10 at 12truept
\font\twelvei=cmmi10 at 12truept
\font\twelveit=cmti10 at 12truept
\font\twelvesl=cmsl10 at 12truept
\font\twelvesy=cmsy10 at 12truept
\font\twelvebf=cmbx10 at 12truept
}

\font\ninei=cmmi9
\font\eighti=cmmi8
\font\sixi=cmmi6
\skewchar\ninei='177 \skewchar\eighti='177 \skewchar\sixi='177

\font\ninesy=cmsy9
\font\eightsy=cmsy8
\font\sixsy=cmsy6
\skewchar\ninesy='60 \skewchar\eightsy='60 \skewchar\sixsy='60

\font\ninebf=cmbx9
\font\eightbf=cmbx8
\font\sixbf=cmbx6

\font\ninett=cmtt9
\font\eighttt=cmtt8

\hyphenchar\tentt=-1 
\hyphenchar\ninett=-1
\hyphenchar\eighttt=-1

\font\ninesl=cmsl9
\font\eightsl=cmsl8

\font\nineit=cmti9
\font\eightit=cmti8


\newskip\ttglue
\def\tenpoint{\def\rm{\fam0\tenrm}%
  \textfont0=\tenrm \scriptfont0=\sevenrm \scriptscriptfont0=\fiverm
  \textfont1=\teni \scriptfont1=\seveni \scriptscriptfont1=\fivei
  \textfont2=\tensy \scriptfont2=\sevensy \scriptscriptfont2=\fivesy
  \textfont3=\tenex \scriptfont3=\tenex \scriptscriptfont3=\tenex
  \def\it{\fam\itfam\tenit}\textfont\itfam=\tenit
  \def\sl{\fam\slfam\tensl}\textfont\slfam=\tensl
  \def\bf{\fam\bffam\tenbf}\textfont\bffam=\tenbf \scriptfont\bffam=\sevenbf
  \scriptscriptfont\bffam=\fivebf
  \normalbaselineskip=12pt
  \let\sc=\eightrm
  \let\big=\tenbig
  \setbox\strutbox=\hbox{\vrule height8.5pt depth3.5pt width\z@}%
  \normalbaselines\rm}

\def\twelvepoint{\def\rm{\fam0\twelverm}%
  \textfont0=\twelverm \scriptfont0=\ninerm \scriptscriptfont0=\sevenrm
  \textfont1=\twelvei \scriptfont1=\ninei \scriptscriptfont1=\seveni
  \textfont2=\twelvesy \scriptfont2=\ninesy \scriptscriptfont2=\sevensy
  \textfont3=\tenex \scriptfont3=\tenex \scriptscriptfont3=\tenex
  \def\it{\fam\itfam\twelveit}\textfont\itfam=\twelveit
  \def\sl{\fam\slfam\twelvesl}\textfont\slfam=\twelvesl
  \def\bf{\fam\bffam\twelvebf}\textfont\bffam=\twelvebf
  \scriptfont\bffam=\ninebf
  \scriptscriptfont\bffam=\sevenbf
  \normalbaselineskip=12pt
  \let\sc=\eightrm
  \let\big=\tenbig
  \setbox\strutbox=\hbox{\vrule height8.5pt depth3.5pt width\z@}%
  \normalbaselines\rm}

\def\fourteenpoint{\def\rm{\fam0\fourteenrm}%
  \textfont0=\fourteenrm \scriptfont0=\tenrm \scriptscriptfont0=\sevenrm
  \textfont1=\fourteeni \scriptfont1=\teni \scriptscriptfont1=\seveni
  \textfont2=\fourteensy \scriptfont2=\tensy \scriptscriptfont2=\sevensy
  \textfont3=\tenex \scriptfont3=\tenex \scriptscriptfont3=\tenex
  \def\it{\fam\itfam\fourteenit}\textfont\itfam=\fourteenit
  \def\sl{\fam\slfam\fourteensl}\textfont\slfam=\fourteensl
  \def\bf{\fam\bffam\fourteenbf}\textfont\bffam=\fourteenbf%
  \scriptfont\bffam=\tenbf
  \scriptscriptfont\bffam=\sevenbf
  \normalbaselineskip=17pt
  \let\sc=\elevenrm
  \let\big=\tenbig
  \setbox\strutbox=\hbox{\vrule height8.5pt depth3.5pt width\z@}%
  \normalbaselines\rm}

\def\seventeenpoint{\def\rm{\fam0\seventeenrm}%
  \textfont0=\seventeenrm \scriptfont0=\fourteenrm \scriptscriptfont0=\tenrm
  \textfont1=\seventeeni \scriptfont1=\fourteeni \scriptscriptfont1=\teni
  \textfont2=\seventeensy \scriptfont2=\fourteensy \scriptscriptfont2=\tensy
  \textfont3=\tenex \scriptfont3=\tenex \scriptscriptfont3=\tenex
  \def\it{\fam\itfam\seventeenit}\textfont\itfam=\seventeenit
  \def\sl{\fam\slfam\seventeensl}\textfont\slfam=\seventeensl
  \def\bf{\fam\bffam\seventeenbf}\textfont\bffam=\seventeenbf%
  \scriptfont\bffam=\fourteenbf
  \scriptscriptfont\bffam=\twelvebf
  \normalbaselineskip=21pt
  \let\sc=\fourteenrm
  \let\big=\tenbig
  \setbox\strutbox=\hbox{\vrule height 12pt depth 6pt width\z@}%
  \normalbaselines\rm}

\def\ninepoint{\def\rm{\fam0\ninerm}%
  \textfont0=\ninerm \scriptfont0=\sixrm \scriptscriptfont0=\fiverm
  \textfont1=\ninei \scriptfont1=\sixi \scriptscriptfont1=\fivei
  \textfont2=\ninesy \scriptfont2=\sixsy \scriptscriptfont2=\fivesy
  \textfont3=\tenex \scriptfont3=\tenex \scriptscriptfont3=\tenex
  \def\it{\fam\itfam\nineit}\textfont\itfam=\nineit
  \def\sl{\fam\slfam\ninesl}\textfont\slfam=\ninesl
  \def\bf{\fam\bffam\ninebf}\textfont\bffam=\ninebf \scriptfont\bffam=\sixbf
  \scriptscriptfont\bffam=\fivebf
  \normalbaselineskip=11pt
  \let\sc=\sevenrm
  \let\big=\ninebig
  \setbox\strutbox=\hbox{\vrule height8pt depth3pt width\z@}%
  \normalbaselines\rm}

\def\eightpoint{\def\rm{\fam0\eightrm}%
  \textfont0=\eightrm \scriptfont0=\sixrm \scriptscriptfont0=\fiverm%
  \textfont1=\eighti \scriptfont1=\sixi \scriptscriptfont1=\fivei%
  \textfont2=\eightsy \scriptfont2=\sixsy \scriptscriptfont2=\fivesy%
  \textfont3=\tenex \scriptfont3=\tenex \scriptscriptfont3=\tenex%
  \def\it{\fam\itfam\eightit}\textfont\itfam=\eightit%
  \def\sl{\fam\slfam\eightsl}\textfont\slfam=\eightsl%
  \def\bf{\fam\bffam\eightbf}\textfont\bffam=\eightbf \scriptfont\bffam=\sixbf%
  \scriptscriptfont\bffam=\fivebf%
  \normalbaselineskip=9pt%
  \let\sc=\sixrm%
  \let\big=\eightbig%
  \setbox\strutbox=\hbox{\vrule height7pt depth2pt width\z@}%
  \normalbaselines\rm}

\def\tenbig#1{{\hbox{$\left#1\vbox to8.5pt{}\right.\n@space$}}}
\def\ninebig#1{{\hbox{$\textfont0=\tenrm\textfont2=\tensy
  \left#1\vbox to7.25pt{}\right.\n@space$}}}
\def\eightbig#1{{\hbox{$\textfont0=\ninerm\textfont2=\ninesy
  \left#1\vbox to6.5pt{}\right.\n@space$}}}

\def\footnote#1{\edef\@sf{\spacefactor\the\spacefactor}#1\@sf
      \insert\footins\bgroup\eightpoint
      \interlinepenalty100 \let\par=\endgraf
        \leftskip=\z@skip \rightskip=\z@skip
        \splittopskip=10pt plus 1pt minus 1pt \floatingpenalty=20000
        \smallskip\item{#1}\bgroup\strut\aftergroup\@foot\let\next}
\skip\footins=12pt plus 2pt minus 4pt 
\dimen\footins=30pc 

\newinsert\margin
\dimen\margin=\maxdimen
\def\titlefont{\seventeenpoint}
\loadtruetwelvepoint 
\loadtrueseventeenpoint
\catcode`\@=\active
\catcode`@=12  
\catcode`\"=\active

\def\eatOne#1{}
\def\ifundef#1{\expandafter\ifx%
\csname\expandafter\eatOne\string#1\endcsname\relax}
\def\notTrue{\iffalse}\def\isTrue{\iftrue}
\def\ifdef#1{{\ifundef#1%
\aftergroup\notTrue\else\aftergroup\isTrue\fi}}
\def\use#1{\ifundef#1\linemessage{Warning: \string#1 is undefined.}%
{\tt \string#1}\else#1\fi}


\global\newcount\refno \global\refno=1
\newwrite\rfile
\newlinechar=`\^^J
\def\ref#1#2{\the\refno\nref#1{#2}}
\def\nref#1#2{\xdef#1{\the\refno}%
\ifnum\refno=1\immediate\openout\rfile=refs.tmp\fi%
\immediate\write\rfile{\noexpand\item{[\noexpand#1]\ }#2.}%
\global\advance\refno by1}
\def\lref#1#2{\the\refno\xdef#1{\the\refno}%
\ifnum\refno=1\immediate\openout\rfile=refs.tmp\fi%
\immediate\write\rfile{\noexpand\item{[\noexpand#1]\ }#2\semi}%
\global\advance\refno by1}
\def\cref#1{\immediate\write\rfile{#1\semi}}

\def\semi{;\hfil\noexpand\break}

\def\inputAuxIfPresent#1{\immediate\openin1=#1
\ifeof1\message{No file \auxfileName; I'll create one.
}\else\closein1\relax\input\auxfileName\fi%
}

\newif\ifWritingAuxFile
\newwrite\auxfile
\def\SetUpAuxFile{%
\xdef\auxfileName{\jobname.aux}%
\inputAuxIfPresent{\auxfileName}%
\WritingAuxFiletrue%
\immediate\openout\auxfile=\auxfileName}


\def\bye{\par\vfill\supereject%
\ifAnyCounterChanged\linemessage{
Some counters have changed.  Re-run tex to fix them up.}\fi%
\end}

\def\ref#1#2{\nref#1{#2}}
\overfullrule 0pt
\hfuzz 52pt
\hsize 6.25 truein
\vsize 8.5 truein

\loadfourteenpoint

\def\c{\,\cdot\,}

\def\eps{\epsilon}

\def\frac#1#2{{#1\over#2}}

\ref\StringBased{Z. Bern and D.A.\ Kosower, Phys.\ Rev.\ Lett.\ 66:1669
(1991)\semi
Z. Bern and D.A.\ Kosower, Nucl.\ Phys. {B379}:451 (1992)\semi
Z. Bern and D.A.\ Kosower, in {\it Proceedings of the PASCOS-91
Symposium}, eds.\ P. Nath and S. Reucroft\semi
Z. Bern, Phys.\ Lett.\ 296B:85 (1992)\semi
Z. Bern, L. Dixon and D.A.\ Kosower, unpublished}

\ref\Tasi{Z. Bern, UCLA/93/TEP/5, hep-ph/9304249, proceedings of TASI 1992}

\ref\FiveGluon{Z. Bern, L. Dixon and D.A.\ Kosower,
Phys.\ Rev.\ Lett. 70:2677 (1993); in preparation}

\ref\Integrals{Z. Bern, L. Dixon and D.A.\ Kosower, Phys.\ Lett.\
B302:299 (1993); SLAC-PUB-5947, to appear in Nucl.\ Phys.\ B}

\ref\Gravity{Z. Bern, D.C. Dunbar, and T. Shimada, Phys.\ Lett.\
B312:277 (1993)}

\ref\Green{%
M.B.\ Green, J.H.\ Schwarz and L. Brink, Nucl.\ Phys.\ B198:472 (1982)}

\ref\Susy{
M.T.\ Grisaru, H.N.\ Pendleton and P.\ van Nieuwenhuizen,
Phys. Rev. {D15}:996 (1977)\semi
M.T. Grisaru and H.N. Pendleton, Nucl.\ Phys.\ B124:81 (1977)\semi
S.J. Parke and T. Taylor, Phys.\ Lett.\ B157:81 (1985)\semi
Z. Kunszt, Nucl.\ Phys.\ B271:333 (1986)\semi
M.L.\ Mangano and S.J. Parke, Phys.\ Rep.\ {200}:301 (1991)}

\ref\Mapping{Z. Bern and D.C.\ Dunbar, Nucl.\ Phys. {B379}:562 (1992)}

\ref\Others{M.J.\ Strassler,  Nucl.\ Phys.\ B385:145 (1992)\semi
M.G.\ Schmidt and C. Schubert, preprint HD-THEP-93-24, hep-th/9309055}

\ref\ZToGammas{M.\ Baillargeon and F. Boudjema, Phys.\ Lett.\ B272:158
(1991)\semi
X.Y. Pham, Phys.\ Lett.\ B272:373 (1991)\semi
F.-X.\ Dong, X.-D. Jiang and X.-J. Zhou, Phys.\ Rev.\ D46:5074 (1992)}

\ref\Morgan{E.W.N.\ Glover and A.G.\ Morgan, Z.\ Phys.\ C60:175 (1993)}

\ref\Chanowitz{M. Chanowitz, Phys.\ Rev.\ Lett.\ 69:2037 (1992)\semi
G.V. Jikia, Phys.\ Lett.\ B298:224 (1993); Nucl.\ Phys.\ B405:24 (1993)\semi
B. Bajc, Phys.\ Rev. D48:1907 (1993)\semi
M.S.\ Berger, preprint MAD/PH/771\semi
D. A.\ Dicus and C. Kao, FSU-HEP-930808, hep-ph/9308330\semi
A. Abbasabadi, D. Bowser-Chao, D.A.\ Dicus and W.W.\ Repko, MSUTH-92-03,
hep-ph/9301226}

\ref\Color{
J.E.\ Paton and H.M.\ Chan, Nucl.\ Phys.\ B10:516 (1969)\semi
F.A.\ Berends and W.T.\ Giele,
Nucl.\ Phys.\ B294:700 (1987)\semi
M.\ Mangano and S.J.\ Parke, Nucl.\ Phys.
B299:673 (1988)\semi
M.\ Mangano, Nucl.\ Phys.\ B309:461 (1988)\semi
Z. Bern and D.A.\ Kosower, Nucl.\ Phys.\ B362:389 (1991)}

\ref\Background{G. 't Hooft,
Acta Universitatis Wratislavensis no.\
38, 12th Winter School of Theoretical Physics in Karpacz; {\it
Functional and Probabilistic Methods in Quantum Field Theory},
Vol. 1 (1975)\semi
B.S.\ DeWitt, in {\it Quantum Gravity II}, eds. C. Isham, R.\ Penrose and
D.\ Sciama (Oxford, 1981)\semi
L.F.\ Abbott, Nucl.\ Phys.\ B185:189 (1981)\semi
L.F\ Abbott, M.T.\ Grisaru and R.K.\ Schaeffer,
Nucl.\ Phys. {B229}:372 (1983)}

\ref\GN{J.L.\ Gervais and A. Neveu, Nucl.\ Phys.\ B46:381 (1972)\semi
        D.A. Kosower, Nucl.\ Phys.\ B335:23 (1990)}

\ref\Tensor{R.\ Karplus and M.\ Neuman, Phys.\ Rev. 80:380 (1950)}

\ref\SpinorHelicity{%
F.\ A.\ Berends, R.\ Kleiss, P.\ De Causmaecker, R.\ Gastmans and T.\ T.\ Wu,
        Phys.\ Lett.\ 103B:124 (1981)\semi
P.\ De Causmaeker, R.\ Gastmans,  W.\ Troost and  T.\ T.\ Wu,
Nucl. Phys. B206:53 (1982)\semi
R.\ Kleiss and W.\ J.\ Stirling,
   Nucl.\ Phys.\ B262:235 (1985)\semi
   J.\ F.\ Gunion and Z.\ Kunszt, Phys.\ Lett.\ 161B:333 (1985)\semi
 R.\ Gastmans and T.T.\ Wu,
{\it The Ubiquitous Photon: Helicity Method for QED and QCD} (Clarendon Press)
(1990)\semi
Z.\ Xu, D.-H.\ Zhang and L. Chang, Nucl.\ Phys.\ B291:392 (1987)}

\ref\Passarino{G. Passarino and M. Veltman, Nucl.\ Phys.\ B{160:151 (1979)}}

\ref\Lam{J.D. Bjorken, Stanford Ph.D. thesis (1958)\semi
J.D. Bjorken and S.D. Drell, {\it Relativistic Quantum Fields}
(McGraw-Hill, 1965)\semi
J. Mathews, Phys.\ Rev. 113:381 (1959)\semi
S. Coleman and R. Norton, Nuovo Cimento 38:438 (1965)\semi
C.S. Lam and J.P. Lebrun, Nuovo Cimento 59A:397 (1969)\semi
C.S. Lam,  Nucl.\ Phys.\ B397:143 (1993)}

\ref\NonLinear{K. Fujikawa, Phys.\ Rev. D7:393 (1973)\semi
M. Base and N.D. Hari Dass, Ann.\ Phys.\ 94:349 (1975)\semi
M.B.\ Gavela, G. Girardi, C. Malleville and P. Sorba,
Nucl.\ Phys.\ B193:257 (1981)\semi
N.G. Deshpande and M. Nazerimonfared, Nucl.\ Phys.\ B213:390 (1983)\semi
F. Boudjema, Phys.\ Lett.\ B187:362 (1987)}

\ref\HV{G. 't\ Hooft and M. Veltman, Nucl.\ Phys.\ B44:189 (1972)}

\ref\Siegel{W. Siegel, Phys.\ Lett.\ 84B:193 (1979)}

\ref\ScalarBox{G. 't Hooft and M. Veltman, Nucl.\ Phys.\ B153:365 (1979)\semi
A. Denner, U. Nierste, and R. Scharf, Nucl.\ Phys. B367:637 (1991)\semi
A. Davydychev and N.Ussyukina, Phys.\ Lett.\ B298 (1993) 363}

\ref\Kunszt{Z. Kunszt, A. Signer and Z. Trocsanyi, preprint  ETH-TH/93-11,
 hep-ph/9305239, to appear in Nucl.\ Phys.\ B}

\ref\Future{Z. Bern, L. Dixon, and D.A.\ Kosower, in preparation}


\hfill UCLA/93/TEP/36

\hfill DTP/93/80

\hfill hep-ph/9312218

\hfill October 1993

\vskip 1 cm

\centerline{\titlefont Supersymmetry Relations Between Contributions To}
\vskip .4 cm

\centerline{\titlefont One-Loop Gauge Boson Amplitudes}

\vskip 2 cm
\centerline{Z. Bern}
\centerline{\it Department of Physics}
\centerline{\it UCLA}
\centerline{\it Los Angeles, CA 90024}
\vskip .3 cm
\centerline{\rm and}
\vskip .3 cm
\centerline{A.G.\ Morgan}
\centerline{\it Department of Physics}
\centerline{\it University of Durham}
\centerline{\it Durham DH1 3LE, England}

\vskip 1.2 truecm \baselineskip12pt


\vskip .5 cm
\centerline{\bf Abstract }

\vskip 0.3 truecm
{
\narrower\smallskip
We apply ideas motivated by string theory to improve the calculational
efficiency of one-loop weak interaction processes with massive
external gauge bosons.  In certain cases ``supersymmetry'' relations
between diagrams with a fermion loop and with a gauge boson loop hold.
This is explicitly illustrated for a particular one-loop standard
model process with four-external gauge bosons.  The supersymmetry
relations can be used to provide further significant improvements in
calculational efficiency.
\smallskip}

\baselineskip15pt

\vfill\break

\vskip .4 truecm
\noindent
{\bf 1. Introduction. }

Even the simplest one-loop gauge boson
amplitudes can be rather formidable to compute. Recently an advance in
the calculation of one-loop gauge boson amplitudes has been made based
on string theory [\use\StringBased,\use\Tasi].  Using this technique
the first calculation of the one-loop five-gluon amplitude has been
performed [\use\FiveGluon,\use\Integrals].  As another example,
one-loop graviton scattering calculations
become relatively simple once the corresponding QCD calculations have
been performed [\use\Gravity].

In the case of QCD, the string-based rules have been interpreted
in terms of a particular set of vertices and organizations whose main
feature is that they lead to relatively efficient computations.
As a bonus, the various contributions to the one-loop amplitude
exhibit simple relations between the gluon and fermion contributions
at the level of the integrands.

In the usual Feynman diagram approach the initial lorentz structure of
the various diagrams bear little
resemblance to each other. Each of the different types of Feynman
diagrams are then separately evaluated.  This may be contrasted to
string theory, where the various particle states are treated more
uniformly, making relationships between the various types of
contributions apparent.  In the calculation of the five-gluon
amplitude [\use\FiveGluon], a striking manifestation of this is that
the gluon loop contribution is rather easy to obtain from the fermion
loop contribution since the two calculations are almost identical.
These relations between fermion and boson loop contributions are
connected to the remarkable simplicity of one-loop amplitudes in $N=4$
super-Yang-Mills, which was first pointed out with the aid of string theory
[\use\Green].  Supersymmetry relations are by now a standard tool in
QCD calculations [\use\Susy]. The conventional supersymmetry
relations are between amplitudes with differing numbers of external
fermions. The relations we discuss here are between diagrams with the same
type of external particles but with differing internal particles.

Here we explain how to reorganize one-loop gauge boson amplitudes
involving $W$'s and $Z$'s to mimic the efficient reorganization for
gluons. As an added bonus in certain cases the manifest relations
between gauge boson and fermion loops are preserved.  These relations
can then be used to provide further significant reductions in the
amount of work involved in a computation.
To do this, we will make direct use of the
field theory lessons obtained from string theory
[\use\Mapping,\use\Others].  The approach presented here is helpful
whenever a one-loop diagram contains a non-abelian vertex.

As a particular example, we will discuss the calculation of the
process $Z\rightarrow 3 \gamma$ [\use\ZToGammas,\use\Morgan] (which is
of some interest for compositeness searches). From the results
of a unitary gauge calculation, in ref.~[\use\Morgan] the striking
relationship between the boson and fermion contributions to the
amplitude was already noted.  We will explicitly show how to make use
of this supersymmetry relationship to significantly improve
calculational efficiency for this process.  With the
superstring-motivated reorganization nearly the entire result for the
$W$-loop contributions can be obtained from the fermion loop
contribution. In processes such as $2 \gamma
\rightarrow 2 Z$ [\use\Chanowitz] (which is of some interest for
searches for ultra-heavy fermions at future photon-photon colliders)
there are additional mixed scalar and gauge-boson loops, but one can
still use the supersymmetry relations to significantly reduce the
computational difficulty of the gauge-boson loop contributions.
For processes, with external $W$'s one loses simple supersymmetry
relations due to the flavor changing in the loop, but there are still
significant advantages to the gauge choices which we describe.

In section 2 we review the supersymmetry relations for the diagrams
that appear in one-loop gauge boson scattering calculations and describe
the application to spontaneously broken theories such as the standard
model.  In section 3 we present the calculation of $Z\rightarrow 3 \gamma$
as an explicit example. In section 4 we comment on other processes such as
$2 \gamma \rightarrow 2 Z$ and provide tables containing the
coupling constants for the various vertices.

\vskip .3 cm
\noindent
{\bf 2. $N=4$ supersymmetry relations.}

Although derived from string theory, the string-based organization can
be understood in ordinary field theory [\use\Mapping,\use\Tasi].
Besides the inherent advantage of obtaining simpler diagrams with an
efficient organization, as an added bonus one obtains relations,
connected to the simplicity of $N=4$ super-Yang-Mills amplitudes,
between gauge-boson and fermion loop diagrams.  The use of these $N=4$
supersymmetry relations as a computational tool was pointed out in
ref.~[\use\FiveGluon] for the one-loop five-gluon amplitude.  With the
string-based organization the relations are manifest at the level of
the integrands of diagrams and can be effectively used as a
computational tool to obtain most of the gauge boson loop contribution
from the fermion loop contribution.

Following the discussion of refs.~[\use\Mapping,\use\Tasi] the key
field theory ingredients for obtaining a good fraction of the
gluon amplitude simplifications of the string-based approach are:

\item
{1)} The Feynman rules should be color ordered
[\use\Color,\use\Tasi].  To a large extent this simply
amounts to rewriting the Yang-Mills
structure constants in terms of traces of commutators of fundamental
representation matrices and considering only one color structure at a
time. This concept is useful in QCD because it reduces the number of
diagrams to be considered.

\item
{2)} Background field Feynman gauge [\use\Background] should be used
in calculations where a non-abelian vertex appears in the loop.  This
gauge is used to construct the one particle irreducible diagrams
describing a gauge invariant effective action. The background field
Feynman gauge is advantageous to use because the vertices are simpler
than in the conventional Feynman gauge.  For the $N=4$ supersymmetry
identities to be manifest it is essential for all vertices of the
one-particle irreducible diagrams to be background field gauge
vertices.

\item
{3)} The second order formalism should be used for the vector part (no
$\gamma_5$) of one particle irreducible diagrams with fermion loops
[\use\Mapping].  This formalism amounts to rewriting the
usual Dirac determinant (for a fermion of unit coupling) as
$$
\eqalign{
\det [\; \slash \hskip - .26 cm D + i m ] & =
\bigl\{ \det [(\slash \hskip - .26 cm D - i m)
(\slash \hskip - .26 cm D + i m)] \bigr\}^{1/2} \cr
& = \bigl\{ \det [D^2 - {\textstyle 1\over 2} \sigma^{\mu\nu} F_{\mu\nu}
+ m^2] \bigr\}^{1/2} \; .  \cr}
\eqn\SecondOrder
$$
With this formalism, the fermion loop contributions
are very similar to those of the gauge bosons.  Additionally, there
is considerable overlap with the calculation of ghost or scalar
loop contributions.

\item
{4)} The scattering amplitudes are constructed by sewing trees onto
the one-particle irreducible diagrams.  One can use standard Feynman
gauge for the trees if one desires.  For gluons, a particularly
convenient gauge for the trees is the non-linear Gervais-Neveu gauge
[\use\GN,\use\Mapping] because of the simple vertices.  It is
obviously advantageous to use different gauges for the tree and loop
parts of the computation since one can optimize the gauge choices to
minimize the computations required in the different parts of the
diagrams.  (Although it might seem strange that two different gauge
choices are used for the loop and tree parts of the Feynman diagrams,
in the background field method this has been justified by Abbott,
Grisaru and Schaeffer [\use\Background]).

\item
{5)} With the background field Feynman gauge and second order fermion
formalism for the one-particle irreducible diagrams, virtually the
entire calculation of a gauge boson loop is contained in the fermion
loop calculation.  This can be used to avoid pointless duplication of
significant portions of the calculation.

\item
{6)} Finally, a decomposition into gauge invariant tensors
[\use\Tensor,\use\Morgan] or spinor helicity methods
[\use\SpinorHelicity] can be used. In this paper we use the former method.
With the tensor decomposition
method one can use the usual Passarino-Veltman [\use\Passarino]
technique for performing tensor integrals.  To use the spinor helicity
technique one first performs those spinor simplifications which are
not obstructed by the presence of loop momentum. Then a Feynman
parametrization is performed to eliminate the loop momentum; the
remaining spinor helicity simplifications can then be performed.  (One
can use an electric circuit analogy [\use\Lam] to arrive at the same
integrand if one desires.)  The Feynman parameter integrals can then
be evaluated using the integration method of ref.~[\use\Integrals].

\vskip .3 cm
Here we apply the latter five ideas to weak interactions and
demonstrate that the gain in computational efficiency is quite
significant.  The application of these ideas is straightforward since
it mainly involves using a different set of Feynman rules than the
conventional ones and then observing a set of relationships
between the integrands of certain diagrams.  In the string-based
approach of refs.~[\use\StringBased,\use\Mapping] these relations are
an inherent property of the string-based rules.  In the above field
theory approach, the relations are found after the trace over
$\gamma$-matrices has been performed and the integrands of the various
loop contributions are compared.  We now present the application
of the above ideas to weak interactions.

First consider the case of no fermions.
In the background field Feynman gauge [\use\Background]
this sector of the
$SU(2) \times U(1)$ Lagrangian is given by ${\cal L}_1 + {\cal L}_2 +
{\cal L}_{gf} + {\cal L}_{ghost}$ where,
$$
\eqalign{
{\cal L}_1 &= -{1 \over 4} \bigl( F_i^{\mu\nu} (\tilde W + W) \bigr)^2
	-{1 \over 4} \bigl( F^{\mu\nu} (\tilde B + B) \bigr)^2
\cr
{\cal L}_2 &= (D_\mu \phi)^\dagger (D^\mu \phi)
	- \lambda (\phi^\dagger \phi)^2 + \mu^2 \phi^\dagger \phi
\cr
{\cal L}_{gf} &= -{1 \over 2 } \bigl( \partial_\mu W^{i\,\mu}
	+ g   \epsilon^{ijk} \tilde{W}_{j\,\mu} W_k^\mu
	+ {i g \over 2} (
		\phi'^\dagger T^i \phi_0
		- \phi_0^\dagger T^i \phi' ) \bigr)^2 \cr
	&-{1 \over 2 } \bigl( \partial_\mu B^{\mu}
	+ {i g' \over 2} (
		\phi'^\dagger \phi_0
		- \phi_0^\dagger \phi' ) \bigr)^2 \cr
{\cal L}_{ghost} &=
       - \omega_i^\dagger \biggl(
                \partial^2 \delta^{il}
                - g {\buildrel \leftarrow \over \partial}_\mu
                         \epsilon^{ijl} (W_j^\mu+\tilde{W}_j^\mu)
                + g \epsilon^{ijl} \tilde{W}_j^\mu
                        {\buildrel \rightarrow \over \partial}_\mu
                + g^2 \tilde{W}_j^\mu (W_m^\mu+\tilde{W}_m^\mu)
                        \epsilon^{ijk} \epsilon^{kml}  \cr
\null & \hskip 1 cm
                + {g^2\over 4} (
                     \phi^\dagger T^l T^i \phi_0 + \phi_0^\dagger T^i T^l \phi)
        \biggr) \omega_l
       - b^\dagger \biggl(
                \partial^2
                + {g'^2\over 4} ( \phi^\dagger \phi_0 + \phi_0^\dagger \phi )
        \biggr) b \cr
\null & \hskip 1cm
       - \omega_i {g g'\over 4}
                ( \phi^\dagger T^i \phi_0 + \phi_0 T^i \phi ) b
       - b^\dagger {g g' \over 4}
                ( \phi^\dagger T^l \phi_0 + \phi_0^\dagger T^l \phi ) \omega_l
}
\anoneqn
$$
where $T^i$ are the Pauli spin matrices and $\tilde W^i$ and $\tilde
B$ are respectively the $SU(2)$ and $U(1)$ hypercharge background
fields and $W^i$ and $B$ are the corresponding quantum fields.  The
covariant derivatives appearing in ${\cal L}_2$ are covariant with
respect to both quantum and background fields.  The ghost
Lagrangian may be obtained by the usual Faddeev-Popov technique.   In order
to obtain the usual fields of the standard model we shift the Higgs
field $\phi = \phi_0 + \phi'$ with
$$
\phi_0 = \left( \eqalign{ &0 \cr v/&\sqrt{2} \cr } \right)
\hskip50pt
\phi' = \left( \eqalign{& \hskip 1 cm \phi^+  \cr
                        & (H + i \chi )/\sqrt{2} \cr } \right)
\anoneqn
$$
with $v = (\mu^2/\lambda)^{1 \over 2}$ and define
$$
\eqalign{
\tilde W_\mu^1 &= (\tilde W^+ + \tilde W^-)\;/\sqrt{2} \cr
\tilde W_\mu^2 &= (\tilde W^+ - \tilde W^-)\;i /\sqrt{2} \cr
\tilde W^3_\mu &= \cos \theta_W \tilde Z_\mu + \sin \theta_W \tilde A_\mu \cr
\tilde B_\mu &= -\sin \theta_W \tilde Z_\mu + \cos\theta_W \tilde A_\mu \cr}
\eqn\ZWDef
$$
with similar equations for the quantum and ghost fields.

After performing the shifts of field variables in eqs.~(2) and (3)
we obtain the gauge sector of the standard model Lagrangian in
background field Feynman gauge.  The Feynman rules generated by this
Lagrangian relevant for the calculation of $Z\rightarrow 3 \gamma$ are
depicted in \fig\FeynmanAFigure.  Only those vertices with two quantum
fields attached are given since those are the only contributing ones
at one loop.  These Feynman rules satisfy the property that there is
no $\tilde A \phi W^\pm $ coupling, considerably reducing the number
of diagrams which must be considered in the $Z\rightarrow 3 \gamma$
calculation, since diagrams with mixed $\phi$-$W$
loops do not appear. (This is similar to the absence
of such couplings in the non-linear $R_\xi$ gauges discussed in
refs.~[\use\NonLinear].)  For generality the coupling constants in the
rules of fig.~\use\FeynmanAFigure\ have been removed since the various
types of gauge bosons couple with different strengths.  The various
coupling constants required for the calculation of $Z\rightarrow 3
\gamma$ are given in Table~1.
\vskip .4 cm
\hskip .01 cm
\hbox{
\def\tend{\cr \noalign{\hrule}}
\def\t#1{\tilde{#1}}
\def\d#1{{#1}^\dagger}
\def\wp{{\omega^+}}
\def\wm{{\omega^-}}
\def\wpm{{\omega^\pm}}
\def\wmp{{\omega^\mp}}
\def\wz{{\omega^Z}}
\def\wa{{\omega^A}}

\def\tw{\theta_W}

\vbox{\offinterlineskip
{
\hrule
\halign{
        &\vrule#
        &\strut\quad#\hfil\vrule
        &\quad\hfil\strut#\hfil\vrule
        \cr
&{\bf Vertex}                   &{\bf Coefficient}      &\tend
%
%
height10pt&$\t{A}W^-W^+$                        &$e$                    &\tend
height10pt&$\t{Z}W^-W^+$                        &$e/\tan\tw$            &\tend
height10pt&$\t{A}\phi^-\phi^+$, $\t{A}\d\wp\wp$, $\t{A}\wm\d\wm$
                                &$-e$                   &\tend
height10pt&$\t{Z}\phi^-\phi^+$          &$-e/\tan2\tw$          &\tend
height10pt&$\t{Z}\d\wp\wp$, $\t{Z}\wm\d\wm$
                                &$-e/\tan\tw$           &\tend
height10pt&$\t{A}\t{A}\phi^+\phi^-$, $\t{A}\t{A}\d\wpm\wpm$
                                &$e^2$                  &\tend
height10pt&$\t{A}\t{Z}\phi^+\phi^-$     &$e^2/\tan2\tw$         &\tend
height10pt&$\t{A}\t{Z}\d\wpm\wpm$               &$e^2/\tan\tw$          &\tend
height10pt&$\t{A}\t{A}W^-W^+$,          &$e^2$                  &\tend
height10pt&$\t{A}\t{Z}W^-W^+$,          &$e^2/\tan\tw$          &\tend
}
}}
}
\nobreak
{\baselineskip 10 pt\narrower\smallskip\noindent\ninerm
{\ninebf Table 1:} The coupling constants of the vertices needed for
the calculation of $Z \rightarrow 3 \gamma$.}

\vskip .3 cm

Now consider the inclusion of internal fermions with no flavor
changing in the loop.  Because the relationship between the fermion
and boson loop that we are interested in does not involve the
$\gamma_5$ in the fermion coupling we divide the fermion loop
computation into a part which contains a $\gamma_5$ and a part which
does not contain a $\gamma_5$. This can be done by considering the
one-particle irreducible diagrams in the conventional (first order)
formalism; one then collects all the $\gamma_5$'s together so that
the fermion trace contains a single $\gamma_5$.  This is then split
into the axial part containing the $\gamma_5$ and the vector part
which does not contain the $\gamma_5$.  The axial part may be
evaluated in the usual way since this part does not play a role in the
supersymmetry identities.  The diagrams of the vector part of the
one-loop effective action may be described by the familiar Dirac
determinant which is rewritten in the second order form
(\use\SecondOrder). It is this form which makes the relationship of the
fermion loop to the gauge boson loop manifest in the integrands.  For
the case where there is flavor changing within the loop, and
necessarily different masses appear inside it, the relationship to the
gauge boson loop is more obscure and one loses the added bonus of
simple supersymmetry relations; the advantage of simpler background
field vertices is, of course, not lost.

In particular, for the case of $Z \rightarrow 3 \gamma$ the $\gamma_5$
contributions all drop out because of cancellations between diagrams where
the fermion circulates in one direction and diagrams where the fermion
circulates in the opposite direction.
This means that for this process the entire
fermion loop can be rewritten in the
second order form [\use\Mapping,\use\Tasi]
$$
{\Gamma}_{\rm fermion}[\tilde W^i, \tilde B] =
{1\over 2}\ln \det [D^2 (\tilde W^i,\tilde B)
- {\textstyle {1 \over 2}}\sigma^{\mu\nu} {\bf F}_{\mu\nu}(\tilde W^i)
- {\textstyle {1 \over 2}}\sigma^{\mu\nu} F_{\mu\nu}(\tilde B) + m^2]
\eqn\OneLoopFermionAction
$$
where the replacement in eq.~(\use\ZWDef) should be carried out to
obtain the usual fields of the standard model.  By expanding out this
determinant we obtain the one-loop Feynman rules for internal fermions
depicted in \fig\FeynmanBFigure. The coupling associated with each
background field is the same as the appropriate effective vector
coupling of the first order formalism with an accompanying loop factor
$-1/2$, where the minus sign is the familiar one for a fermion
loop. One obvious feature of these second order fermion rules is that
they bear a much greater resemblance to the boson rules than the
conventional (first order formalism) Feynman rules for fermions; this
is important for making the supersymmetry relations hold diagram-by-diagram.

With the rules given in figs.~\use\FeynmanAFigure\ and
\use\FeynmanBFigure\ the {\it integrands} of diagrams for one-loop $n$ gauge
boson scattering satisfy an $N=4$ supersymmetry constraint
[\use\FiveGluon,\use\Tasi]. This relationship between diagrams with
fermions in the loop and gauge bosons (and associated ghosts)
in the loop is depicted in \fig\SusyFigure\ and is given by
$$
\eqalign{
&D^{\rm scalar}(m_s) =  C_s S (m_s) \cr
&D^{\rm fermion}(m_{\! f}) = - C_f (2 S (m_{\! f}) +
 F (m_{\! f})) \cr
&D^{\rm gauge\ boson}(m_g) = C_g
((1 - \delta_R \eps) 2 {\cal S}(m_g)
+ 4 F(m_g) + G (m_g))\cr }
\eqn\SusyIdent
$$
where the particle labels refer to the states circulating in the loop,
the $m_x$ are the masses of the particles circulating in the loop and
the $C_i$ are coupling constant factors which depend on the processes
under consideration.  The $D$ all refer to the same diagram types,
but with different particles circulating in the loops.  For
two or three legs attached to the loop the simple quantity $G$ vanishes
at the level of the integrand.
(The dimensional
regularization parameter is $\delta_R=1$ for either conventional
dimensional regularization or for the 't~Hooft-Veltman scheme
[\use\HV] while $\delta_R=0$ for either
the dimensional reduction  [\use\Siegel] or
four-dimensional helicity [\use\StringBased] schemes.)
In cases where all types of diagrams satisfy the supersymmetry identities
(\use\SusyIdent) (such as $Z\rightarrow 3
\gamma$), the sum over all diagrams -- namely the amplitude -- also
satisfies this identity.

The connection of these identities to $N=4$ supersymmetry is that for
$N=4$ super-Yang-Mills (one gluon, four Weyl
fermions, and 6 real scalars) everything but $G$ cancels after summing
over the various loop contributions. (The regulator factor
$\delta_R=0$ is necessary so that supersymmetry is not broken).  That is,
$$
D^{N=4\ \rm susy} = g^4 G
\anoneqn
$$
where $g$ is the coupling. The other terms all cancel.

In performing the calculation, instead of calculating the diagrams
directly it is more efficient to calculate $S$, $F$ and $G$.  The
importance of the above identities is that each part of the
calculation is successively easier to perform; $S$ is the most
complicated part, $F$ is the next most complicated part and $G$ is by
far the easiest part of the calculation. In a conventional
approach one would effectively be recomputing the $S$ and $F$ parts
since one computes the gauge boson loop directly.  This leads to a
significant computational advantage for the gauge boson loop beyond
the already large simplifications of Feynman background field gauge.
(With conventional gauge choices, like 't~Hooft-Feynman
or unitary gauge, the unnecessary recomputation of $S$ and $F$ is
actually significantly more complicated than the direct computation of
these quantities from scalar and fermion loops.)

\vskip .3 truecm
\noindent
{\bf 3. Explicit example. }

Consider the process $Z\rightarrow 3 \gamma$. This process has already
been discussed in refs.~[\use\ZToGammas,\use\Morgan] using more conventional
techniques. We show here how
to reduce the $W$-loop computation to a very simple one once
the fermion loop is calculated.  The four one-loop diagram types required for
calculating $Z\rightarrow 3 \gamma$ are depicted in
\fig\ZToGammasFigure.  The complete amplitude is obtained by summing
over the six permutations of external legs.

{}From ref.~[\use\Morgan] we have the general tensor consistent with
gauge invariance and crossing symmetry for the three photons as
$$
\eqalign{
{\cal M}^{\alpha \mu \nu \rho}(p_1,p_2,p_3) =
&A_1(p_1,p_2,p_3)~\frac{1}{p_1\c p_3} ~
\left (\frac{p_3^\mu p_1^\rho}{p_1\c p_3} -  g^{\mu \rho} \right) p_1^\alpha
 \left (\frac{p_3^\nu}{p_2\c p_3} -  \frac{p_1^\nu}{p_1\c p_2} \right) \cr
& + A_2(p_1,p_2,p_3) ~\left\{ \frac{1}{p_2\c p_3}
\left(\frac{p_1^\alpha p_3^\mu}{ p_1\c p_3} - g^{\alpha \mu}\right )
\left(\frac{p_1^\nu  p_2^\rho}{p_1\c p_2} - g^{\nu \rho}\right )\right. \cr
&\hskip2.2cm+ \left. ~\frac{1}{p_1\c p_3}
\left ( \frac{p_1^\nu}{p_1\c p_2}- \frac{p_3^\nu}{p_2\c p_3} \right)
\left (p_1^\rho g^{\alpha \mu} -p_1^\alpha g^{\mu\rho} \right ) \right\} \cr
&+A_3(p_1,p_2,p_3) ~\frac{1}{p_1\c p_3}  ~
\left (\frac{p_1^\alpha p_3^\mu}{p_1\c p_3} -g^{\alpha \mu} \right)
\left (\frac{p_3^\nu p_2^\rho}{p_2\c p_3} - g^{\nu \rho} \right). \cr }
\eqn\KinTensor
$$
The amplitude is obtained from this tensor by dotting it into the
external polarization vectors.  In this method one only computes
the scalar quantities $A_1$, $A_2$, $A_3$ thereby eliminating the
redundant information contained in a gauge invariant expression, in a
way analogous to what happens with
spinor helicity methods. Factoring out the coupling
constants we obtain an expression for the $A_i$'s in terms of the
scalar, fermionic and gauge boson Feynman rules of
figs.~\use\FeynmanAFigure\ and ~\use\FeynmanBFigure\ (c.f. (12) of
[\use\Morgan])
$$
\eqalign{
A_i(p_1,p_2,p_3) &= {i e^4 \over 8\pi^2\sin 2\theta} \biggl(
	\sum_f e_f^3 v_f A_i^f(s,t,m_f)
\cr&
+	\cos^2 \theta_W  A_i^W(s,t,M_W)
+	{\cos 2 \theta_W \over 2} A_i^\phi(s,t,M_W)
\biggr)
}
\eqn\ZpppAmplitude
$$
where the fermionic $A_i$'s are defined to include their overall
minus sign. For the gauge loop we note that the inclusion of both
ghosts, $\omega^\pm$ and $\omega^\pm{}^\dagger$
is straightforward since they are
just (fermionic) complex scalar fields. In fact, in background field
gauge, $A_i^{\omega^\pm}(s,t,M_W) =
- A_i^\phi(s,t,M_W)$, and thus $A_i^W$,
which we take to include both the $W$ and Faddeev-Popov ghost contributions,
is obtained by application of the Feynman rules of
fig.~\use\FeynmanAFigure\ minus twice the scalar $A_i^\phi$ result.  We
therefore only need to compute the three separate contributions $A_i^\phi$,
$A_i^f$ and $A_i^W$.  Further discussion of the tensor decomposition
method can be found in refs.~[\use\Tensor,\use\Morgan].

In order to minimize the duplication of effort we make use of the
supersymmetry relations (\use\SusyIdent) to systematize our evaluation
of the above scalar $A_i$ functions. Since all of the diagram types in
this calculation satisfy the supersymmetry relation (\use\SusyIdent)
the sum over the diagrams or amplitude will satisfy the relation.
As mentioned previously, it
is not difficult to verify that the $\gamma_5$ contribution in the
fermion loop drops out because of cancellations between diagrams where
the fermion circulates in one direction and diagrams where the fermion
circulates in the opposite direction.  This means that the entire fermion
loop contribution is of the vector type and therefore
included in the supersymmetry identity.

The first step is to compute the
scalar loop contribution. After summing over diagrams and reducing the
tensor integral down to scalar ones [\use\Passarino] the result is
$$
\eqalign{
A_1^\phi(s,t,m) & = {\cal S}_1 (s,t,m) =
 -{1\over 2} A_1^f(s,t, m),
\cr 
 A_2^\phi(s,t,m) & = {\cal S}_2 (s,t,m) = -{1\over 2} A_2^f(s,t, m) -
           {tu \over 2} \Bigl(2 m^2 H(m) - {1\over s} E(t,u,m) \Bigr),
\cr 
A_3^\phi(s,t,m) & = {\cal S}_3 (s,t,m) =  -{1\over 2} A_3^f(s,t, m) -
{1\over 2 t} \Bigl\{ t^2 E(s,t,m) - u^2 E(u,s,m)
\cr
\null & \hskip 1.5 cm + 2 m^2 ut(u-t) H(m) \Bigr\} \cr} 
\anoneqn
$$
where the $A_i^f$ are functions defined in ref.~[\use\Morgan] and
$$
\eqalign{
& E(s,t,m) = s\; C(s,m)+ s_1 C_1(s,m) +t\; C(t,m) +t_1 C_1(t,m)
- st\; D(s,t,m), \cr
& H(m) = D(s,t,m) + D(t,u,m) + D(u,s,m) \cr
& s_1 C_1(s,m) = s C(s,m) - M_Z^2 C(M_Z^2,m) \cr
}
\anoneqn
$$
The mass $m$ is the mass of the scalar going around the loop and
$s=(p_1 +p_2)^2$, $t=(p_1+p_4)^2$ and $u=(p_1 +p_3)^2$ are Mandelstam
variables, and $s_1 = s - M_Z^2$, $t_1 = t - M_Z^2$ and $u_1 = u - M_Z^2$.
For convenience we quote the simplest of these functions here as
$$
\eqalign{
{\cal S}_1(s,t,m)  &=
-\frac{2st}{t_1} - \frac{4t}{u}\biggl( s ~B_1(s,m) - s_1 ~B_1(t,m)\biggr)
+\frac{2M_Z^2(s+2u)t}{t_1^2}~B_1(t,m) \cr
&-\frac{st(2t+u)}{u^2}~E(s,t,m) - \frac{4m^2t}{u}~E(s,t,m)
- \frac{2m^2t}{s}~E(t,u,m) \cr
- & 2m^2 \biggl (s~C(s,m)+ t~C(t,m) + u_1~C_1(u,m) \biggr)\cr
+ & \frac{4m^2(s+2u)t}{t_1}~C_1(t,m) +\frac{2m^2s t (u+2t)}{u} ~D(s,t,m)\cr
&+ m^2 \biggl( u t~D(t,u,m) + s t~D(s,t,m) + u s ~D(u,s,m) \biggr )
+4m^4 t~H(m) \cr }
\eqn\MorganFunc
$$
with $B_1(s,m) = B(s,m) - B(M_Z^2,m)$.
The scalar loop is the most complicated piece to integrate
since the graphs contain the most powers of loop momentum
in the numerator.  In general, because of the explosion of terms which
occurs in the evaluation of tensor integrals
[\use\Passarino,\use\Integrals], factors of loop momenta cause the
largest complications; this is reflected in the complexity of
this result.

The next stage of the computation is to subtract out the part of the
fermion loop proportional to the scalar loop in the integrand of each
diagram; after integration this yields the ${\cal F}_i$ which after
summing over diagrams are as follows
$$
\eqalign{
{\cal F}_1 (s,t,m) & = - A_1^f (s,t,m) - 2 {\cal S}_1 (s,t,m) = 0,  \cr
{\cal F}_2 (s,t,m) & = - A_2^f (s,t,m) - 2 {\cal S}_2 (s,t,m) =
          tu \Bigl(2 m^2 H(m) - {1\over s} E(t,u,m) \Bigr), \cr
{\cal F}_3 (s,t,m) & = - A_3^f (s,t,m) - 2 {\cal S}_3 (s,t,m) =
{1\over t} \Bigl\{ t^2 E(s,t,m) - u^2 E(u,s,m)
\cr &
	+ 2 m^2 ut(u-t) H(m) \Bigr\}.  \cr }
\anoneqn
$$
The required integrals are much simpler quantities than for ${\cal S}_i$
since the integrands contain at most two powers of loop momentum
instead of four.  The relative simplicity of the computation as
compared to the scalar loop calculation is reflected in the relative
simplicity of the results.  Plugging the ${\cal S}_i$ and ${\cal F}_i$
into the second of the supersymmetry relations in
eq.~(\use\SusyIdent) reproduces the results for fermion loops of
ref.~[\use\Morgan].

To obtain the $W^{\pm}$ loop contribution in each diagram we subtract
the integrands associated with $2 {\cal S}(s,t, M_W) + 4 {\cal
F}(s,t,M_W)$ from the full expression for the $W^{\pm}$ loop
(including Fadeev-Popov ghosts); this leaves only box diagrams to be
evaluated since all other integrands cancel by the supersymmetry
relations given in fig.~\use\SusyFigure.  Furthermore, the only
terms in the vertices which contribute are those which contain no loop
momentum, the ones containing loop momentum manifestly cancel in the
calculation of ${\cal G}(s,t,M_W)$.  This cancellation is a direct
consequence of the $N=4$ supersymmetry relations.  Since the terms
with loop momentum cancel, $\cal G$ is reduced to
a relatively simple algebraic expression times a scalar box integral,
which may be obtained from ref.~[\use\ScalarBox].
Since there is no need to evaluate
a tensor integral, this part of the computation is relatively trivial.
(Indeed, by using rules of the string based type [\use\StringBased,\use\Tasi]
it is possible to write down the answer without calculation.)
For the diagrams with a 1,2,3,4 and reversed ordering
of legs the remaining kinematic tensor is simple and given by
$$
\eqalign{
 G^{\alpha \mu \nu \rho} (1234) & = - D(s,t) \biggl( 8 \Bigl(
		g^{\alpha \mu} g^{\nu \rho} s u
		+ g^{\alpha \nu} g^{\mu \rho} s t
		+ g^{\alpha \rho} g^{\mu \nu} u t \Bigr) \cr
       & + 16 s \Bigl(
	g^{\alpha \rho} p_3^{\mu} p_3^{\nu}
	+ p_3^{\mu} ( g^{\nu \rho} p_2^{\alpha} - g^{\alpha \nu} p_2^{\rho} )
	+ p_3^{\nu} ( g^{\mu \rho} p_1^{\alpha} - g^{\alpha \mu} p_1^{\rho} )
	\Bigr) \cr
       & + 16 t \Bigl(
	g^{\alpha \mu} p_1^{\nu} p_1^{\rho}
	+ p_1^{\nu} ( g^{\mu \rho} p_3^{\alpha} -  g^{\alpha \rho} p_3^{\mu} )
	+ p_1^{\rho} ( g^{\mu \nu}  p_2^{\alpha} - g^{\alpha \nu}  p_2^{\mu} )
	\Bigr) \cr
       & + 16 u \Bigl(
	g^{\alpha \nu} p_2^{\mu} p_2^{\rho}
	+ p_2^{\rho} ( g^{\mu \nu}  p_1^{\alpha} - g^{\alpha \mu}  p_1^{\nu} )
	+ p_2^{\mu} ( g^{\nu \rho} p_3^{\alpha}  - g^{\alpha \rho} p_3^{\nu} )
	\Bigr) \biggr)
}
\eqn\GKinematicCoeff
$$
where we have organized the terms to exhibit manifest gauge invariance.
The other orderings of external legs are obtained by a relabeling of legs.
After summing over the independent orderings and comparing
to the kinematic tensor (\use\KinTensor) and using
$$
{\cal G}_i (s,t,M_W) = A^W_i (s,t,M_W) - (2
{\cal S}_i (s,t,M_W) + 4 {\cal F}_i(s,t,M_W))
\anoneqn
$$
the result
can be summarized in terms of the three scalar functions
$$
\eqalign{
{\cal G}_1 (s,t,M_W) & = 0,  \cr
{\cal G}_2 (s,t,M_W) & = -2 stu H(M_W),  \cr 
{\cal G}_3 (s,t,M_W) & = 0 \; .  \cr }
\anoneqn
$$
Again the simplicity of the calculation is reflected in the simplicity
of the result. Inserting these functions into the supersymmetry
identities (\use\SusyIdent) reproduces the results of
ref.~[\use\Morgan] for the gauge boson loop. In particular eliminating
for ${\cal S}_i$ in favor of $A^f_i$, and using the identity $\cos
2\theta_W = \cos^2\theta_W ( 2 - M_Z^2/M^2_W )$ we find that the
non-fermionic contribution to the $Z\gamma\gamma\gamma$ scattering
tensor
$$
\eqalign{
A^b_i & = A^W_i + {\cos 2\theta_W \over 2} A^\phi_i \cr
	& = {1 \over 4} \biggl( {M_Z^2 \over M_W^2} - 6 \biggr) A^f_i
	+ {1 \over 4} \biggl( {M_Z^2 \over M_W^2} + 10 \biggr) {\cal F}_i
	+ {\cal G}_i.
}
\anoneqn
$$ This then provides an explanation for the empirically observed
relations (15) of ref.~[\use\Morgan]; namely that they are
supersymmetry identities.

As a simple check on the results for ${\cal G}$, we have verified that
for external mass $M_Z\rightarrow 0$ the kinematic coefficient of
the box diagram given in eq.~(\use\GKinematicCoeff) is
proportional to the color ordered Yang-Mills tree. This is in agreement with
expectations from superstring theory with $N=4$ space-time
supersymmetry [\use\Green] where the one-loop four-point
amplitude is also proportional to the tree.

The calculation we have presented for the $W$ loop may be compared to
the unitary gauge calculation presented in ref.~[\use\Morgan].  In
that paper, the unitary gauge was used because of the significant
reduction in the number of diagrams as compared to the standard
't~Hooft-Feynman gauge. In the string-motivated organization presented
here we have retained all the diagrammatic advantages of the unitary
gauge. In addition, since it has been possible to use a simple Feynman
type background field gauge it has not been necessary at any stage to
cancel superficial ultra-violet divergences arising from the extra
powers of momentum in the unitary gauge propagator. (This was the most
time consuming part of the tensor reduction of ref.~[\use\Morgan]).
Furthermore, the vertices of background field 't Hooft-Feynman gauge
are simpler than those of the unitary gauge. Finally, by making use of
the $N=4$ supersymmetry relations we have reduced the $W$-loop
calculation to that of simple scalar box integrals which are given in
refs.~[\use\ScalarBox]. The reorganization we have presented
therefore represents a clear computational advantage.

What about the fermion loop part of the calculation?  Superficially it
might seem that since there are {\it four} diagram types in the second
order formalism (figs.~\use\FeynmanBFigure\ and
\use\ZToGammasFigure),
instead of a single diagram type in the more usual spinor
based (first order) formalism this represents a retrograde step in the
calculational technique.  In fact, the use of the second order
formalism significantly improves the calculational efficiency of the
fermion loops since most of the calculation can be directly obtained
from the calculation of scalars or ghosts in the loop.  The similarity in
structure of the fermion to scalar vertices
(figs.~\use\FeynmanAFigure\ and
\use\FeynmanBFigure) ensures that when calculating the ${\cal F}_i$
the cancellations between the scalar and fermion loops implied by the
supersymmetry equations (\use\SusyIdent) occur {\it on the first line}
at the level of the integrand and before the evaluation of any tensor
integrals. (Even if one were not interested in scalar or gauge boson
loop contributions, it is generally still advantageous to break the fermion
loop contribution into two separate pieces since it is usually easier
to handle smaller physical pieces in a large calculation.)
The second order formalism therefore also represents a considerable advance in
calculational efficiency for the vector part of fermion loops (with no
flavor changing).

\vskip .3 cm
\noindent
{\bf 4. Other Processes.}

The string-motivated reorganization discussed above is useful for other
amplitudes.  For completeness the coupling constants for the various
other vertices with external gauge bosons are presented in Tables 2-4.
Besides the vertex structures already encountered in \use\FeynmanAFigure\
there is an additional non-abelian vertex given in \fig\NonAbelianFigure ;
the coupling constants associated with this vertex are presented in Table 4.

\vskip 1 cm

\hskip .01 cm
\hbox{
\def\tend{\cr \noalign{\hrule}}
\def\t#1{\tilde{#1}}
\def\d#1{{#1}^\dagger}
\def\wp{{\omega^+}}
\def\wm{{\omega^-}}
\def\wpm{{\omega^\pm}}
\def\wmp{{\omega^\mp}}
\def\wz{{\omega^Z}}
\def\wa{{\omega^A}}

\def\tw{\theta_W}

\vbox{\offinterlineskip
{
\hrule
\halign{
        &\vrule#
        &\strut\quad#\hfil\vrule
        &\quad\hfil\strut#\hfil\vrule
        \cr
&{\bf Vertex}                   &{\bf Coefficient}      &\tend
%
%
height10pt&$\t{W}^+ A W^-$, $\t{W}^- W^+ A$
                                &$e$                    &\tend
height10pt&$\t{W}^+ Z W^-$, $\t{W}^- W^+ Z$
                                &$e/\tan\tw$            &\tend
height10pt&$\t{Z}W^\pm \phi^\mp$        &$-e^2v/\sin\tw\sin2\tw$
                                                        &\tend
height10pt&$\t{Z}ZH$                    &$2e^2v/\sin^22\tw$
                                                        &\tend
height10pt&$\t{W}^\pm W^\mp H$          &$e^2v/2\sin^2\tw$              &\tend
height10pt&$\t{W}^\pm W^\mp \chi$       &$\mp ie^2v/2\sin^2\tw$         &\tend
height10pt&$\t{W}^\pm A\phi^\mp$,       &$e^2v/\sin\tw$                 &\tend
height10pt&$\t{W}^\pm Z\phi^\mp$,       &$e^2v/\tan2\tw\sin\tw$         &\tend
height10pt&$\t{W}^+ H\phi^-$, $\t{W}^- \phi^+ H$
                                &$e/2\sin\tw$           &\tend
height10pt&$\t{W}^\pm\chi\phi^\mp$      &$ie/2\sin\tw$          &\tend
height10pt&$\t{Z}\chi H$                &$-ie/\sin2\tw$         &\tend
}
}}
}
\nobreak
{\baselineskip 10 pt\narrower\smallskip\noindent\ninerm {\ninebf Table
2:} The coupling constants associated with other three point vertices.
Those involving an odd number of gauge fields may be found in
fig.~\use\FeynmanAFigure\ and the remaining vertices involving two
gauge fields are to be found in fig.~\use\NonAbelianFigure.}

\vskip 1.5 cm
\hbox{
\def\tend{\cr \noalign{\hrule}}
\def\t#1{\tilde{#1}}
\def\d#1{{#1}^\dagger}
\def\wp{{\omega^+}}
\def\wm{{\omega^-}}
\def\wpm{{\omega^\pm}}
\def\wmp{{\omega^\mp}}
\def\wz{{\omega^Z}}
\def\wa{{\omega^A}}

\def\tw{\theta_W}

\vbox{\offinterlineskip
{
\hrule
\halign{
        &\vrule#
        &\strut\quad#\hfil\vrule
        &\quad\hfil\strut#\hfil\vrule
        \cr
&{\bf Vertex}                   &{\bf Coefficient}      &\tend
%
%
%
height10pt&$\t{W}^+\t{W}^-AA$
                                &$e^2$                  &\tend
height10pt&$\t{W}^+\t{W}^-AZ$
                                &$e^2/\tan\tw$          &\tend
height10pt&$\t{Z}\t{Z}W^-W^+$, $\t{W}^+\t{W}^-ZZ$
                                &$e^2/\tan^2\tw$        &\tend
height10pt&$\t{W}^\pm\t{W}^\pm W^\mp W^\mp$,
                                &$-e^2/\sin^2\tw$       &\tend
height10pt&$\t{W}^+\t{W}^-\d\wa\wa$     &$e^2$                  &\tend
height10pt&$\t{W}^\pm \t{A}\d\wpm\wa$, $\t{W}^\pm\t{A}\d\wa\wmp$
                                        &$-e^2$                 &\tend
height10pt&$\t{W}^+\t{W}^-\d\wpm\wpm$,  &$e^2/\sin^2\tw$        &\tend
height10pt&$\t{W}^\pm\t{W}^\pm\d\wpm\wmp$,
                                &$-e^2/\sin^2\tw$       &\tend
&\vbox{\hbox{$\t{W}^\pm\t{Z}\d\wpm\wa$, $\t{W}^\pm\t{Z}\d\wa\wmp$}
\hbox{$\t{W}^\pm\t{A}\d\wpm\wz$, $\t{W}^\pm\t{A}\d\wz\wmp$}}
                                &$-e^2/\tan\tw$         &\tend
height10pt&$\t{Z}\t{Z}HH$, $\t{Z}\t{Z}\chi\chi$
                                &$e^2/\sin^22\tw$       &\tend
height10pt&$\t{Z}\t{Z}\phi^+\phi^-$     &$e^2/\tan^22\tw$       &\tend
height10pt&$\t{W}^+\t{W}^-\phi^+\phi^-$, $\t{W}^+\t{W}^-HH$,
$\t{W}^+\t{W}^-\chi\chi$
                                &$e^2/4\sin^2\tw$       &\tend
height10pt&$\t{W}^\pm\t{A}H\phi^\mp$
                                &$e^2/4\sin\tw$         &\tend
height10pt&$\t{W}^\pm\t{Z}H\phi^\mp$
                                &$-e^2/4\cos\tw$        &\tend
height10pt&$\t{W}^\pm\t{A}\chi\phi^\mp$
                                &$\pm ie^2/4\sin\tw$    &\tend
height10pt&$\t{W}^\pm\t{Z}\chi\phi^\mp$
                                &$\mp ie^2/4\cos\tw$    &\tend
height10pt&$\t{W}^+\t{W}^-\d\wz\wa$, $\t{W}^+\t{W}^-\d\wa\wz$
                                &$e^2/\tan\tw$          &\tend
height10pt&$\t{W}^+\t{W}^-\d\wz\wz$, $\t{Z}\t{Z}\d\wpm\wpm$
                                &$e^2/\tan^2\tw$        &\tend
height10pt&$\t{W}^\pm\t{Z}\d\wpm\wz$, $\t{W}^\pm\t{Z}\d\wz\wmp$,
                                &$-e^2/\tan^2\tw$       &\tend
}
}}
}
{\baselineskip 10 pt\narrower\smallskip\noindent\ninerm
{\ninebf Table 3:} The coupling constants associated with the various
four-point vertices found in fig.~\use\FeynmanAFigure.}

\vskip .5 cm
\hskip 1 cm
\hbox{
\def\tend{\cr \noalign{\hrule}}
\def\t#1{\tilde{#1}}
\def\tw{\theta_W}

\vbox{\offinterlineskip
{
\hrule
\halign{
        &\vrule#
        &\strut\quad#\hfil\vrule
        &\quad\hfil\strut#\hfil\vrule
        \cr
&{\bf Vertex}                   &{\bf Coefficient}      &\tend
height10pt&$\t{W}^\pm\t{A}W^\mp A$      &$e^2$                  &\tend
height10pt&$\t{W}^\pm\t{Z}W^\mp A$, $\t{W}^\pm\t{A}W^\mp Z$
                                &$e^2/\tan\tw$          	&\tend
height10pt&$\t{W}^\pm\t{Z}W^\mp Z$
                                &$e^2/\tan^2\tw$        	&\tend
height10pt&$\t{W}^+\t{W}^-W^+W^-$       &$-e^2/\sin^2\tw$       &\tend
}
}}
}
\nobreak
{\baselineskip 10 pt\narrower\smallskip\noindent\ninerm
{\ninebf Table 4:} The coupling constants associated with
four-point vertices of the type in fig.~\use\NonAbelianFigure}

\vskip .3 cm

Using these tables, one could for example consider the
one-loop process $2 \gamma \rightarrow 2 Z$ [\use\Chanowitz] (which is
of some interest to future photon-photon colliders).  In this process
one can again use the $N=4$ supersymmetry relations of
fig.~\use\SusyFigure\ to relate the diagrams with the $W$ going around
the loop to the diagrams with fermions going around the loop.  In this
case, however, there are mixed diagrams with both $W$'s and $\phi$'s
in the loop.  Although such diagrams are apparently not simply related
to fermion loop diagrams they are simpler to evaluate since they have
a maximum of two powers of loop momentum in the numerator.

Due to the simplicity of the background field vertices as well as the
supersymmetry relations (\use\SusyIdent), one can expect a significant
efficiency over previous calculations of $2 \gamma \rightarrow 2 Z$
[\use\Chanowitz].  For example in the one performed by Berger in
standard 't~Hooft-Feynman gauge, there were 188 diagrams to evaluate
for the boson loop contributions. Since each of the vertices is
relatively complicated compared to background field vertices, this
calculation is significantly more complicated than one which follows the
above strategy.  Indeed, Bajc in his paper states that there are
608 terms in the $W$ box diagram alone.
We may also contrast the above strategy to the
non-linear gauge used by Jikia in his calculation; we retain the
advantage of eliminating the $\tilde A \phi^\pm W^{\mp}$ vertex
and have the additional advantages of having simpler vertices and
supersymmetry relations between diagrams.  A third alternative is the
non-linear gauge used by Dicus and Kao which has the advantage of
eliminating all remaining diagrams with mixed $W$-$\phi$ loops, but
then the vertices are more complicated and one loses the supersymmetry
relations for the unmixed diagrams.  Due to the supersymmetry
relations, the main part of each of these calculations only reproduces
pieces already computed for the fermion loops.

The ideas discussed above can also be applied to the case of external
fermions.  In particular, background field Feynman
gauge is still advantageous to use even when some external legs are
fermions.  As for the purely external gauge boson case it is also
useful to identify parts of the calculation which are duplicated in
the various diagrams.  This type of strategy has already been
successfully applied in the calculation of the one-loop corrections to
four- [\use\Kunszt] and five-parton [\use\Future] processes.

\vskip .3 cm
\noindent
{\bf 5. Conclusions. }

Various contributions to gauge boson amplitudes have
relations between them connected to the fact that amplitudes in
$N=4$ super-Yang-Mills have extremely simple forms.  These
relations were first applied in the string-based calculation of
gluon amplitudes [\use\FiveGluon,\use\Tasi].  In order to make
practical use of the supersymmetry relations one needs a formalism
where the relations hold between the integrands of diagrams.
The guidance for constructing such a formalism is provided by string
theory and amounts to special gauge choices and organizations of the
diagrams.

In this paper we have described the supersymmetry relations in weak
interaction processes which involve gauge bosons. These types of
relationships were observed to hold in the explicitly computed weak
interaction process $Z\rightarrow 3 \gamma$ [\use\Morgan], although in
the unitary gauge where the calculation was performed the relations
seem mysterious. We have shown how to reorganize this
calculation as well as other processes so that the supersymmetry
relations are manifest in all stages of the calculation.  Important
ingredients for making the relationships manifest in the diagrams are
the background field Feynman gauge for the gauge-boson loops and the
second order formalism for fermion loops.  In this way the gauge
boson and fermion loop computations have considerable overlap.
The parts of the calculation which overlap do not need to be recomputed
for the gauge boson loop contributions.

A practical consequence of the reorganized calculation and the
manifest supersymmetry relations is that instead of the $W$-loop
contribution being the most complicated part of the calculation it is
relatively easy to obtain it using results from the fermion loop
contribution.

\vskip .2 cm
We thank L.~Dixon, D.C.~Dunbar, E.W.N.~Glover, D.A.~Kosower, and D.
Morris for helpful discussions and G.~Chalmers for collaboration on
early stages of this work.  The work of ZB was supported in part by
the US Department of Energy Grant DE-FG03-91ER40662 and in part by the
Alfred P. Sloan Foundation Grant BR-3222.  AGM would like to thank the
United Kingdom Science and Engineering Research Council for the award
of a Research Studentship, and further to acknowledge the kind
hospitality of the UCLA Physics Department where this work was
completed.

\vfill\eject\immediate\closeout\rfile
\centerline{{\bf References}}\bigskip\frenchspacing%
\input refs.tmp\vfill\eject\nonfrenchspacing

\centerline{\bf Figure Captions}

\vskip .5 cm

\item
{\bf Fig.~\use\FeynmanAFigure:} The vertices, with coupling constants
removed, needed for the calculation of boson loop contributions to
$Z\rightarrow 3 \gamma$.

\item
{\bf Fig.~\use\FeynmanBFigure:} The vertices, with coupling constants
removed, need for the calculation of fermion loop contributions to
$Z\rightarrow 3 \gamma$.

\item
{\bf Fig.~\use\SusyFigure:} The $N=4$ supersymmetry relations.  These
relations hold in the integrands of the diagrams.   (For simplicity the
ghost loop is implicitly included in the gauge boson loop.)

\item
{\bf Fig.~\use\ZToGammasFigure:} The diagrams needed for the calculation
of $Z\rightarrow 3 \gamma$; the loops can be either fermions, gauge bosons
or scalars.

\item
{\bf Fig.~\use\NonAbelianFigure:} The non-abelian three-point vertex
with two gauge bosons appearing in the processes of Table 2 and the
four-point vertex appearing in Table 4.

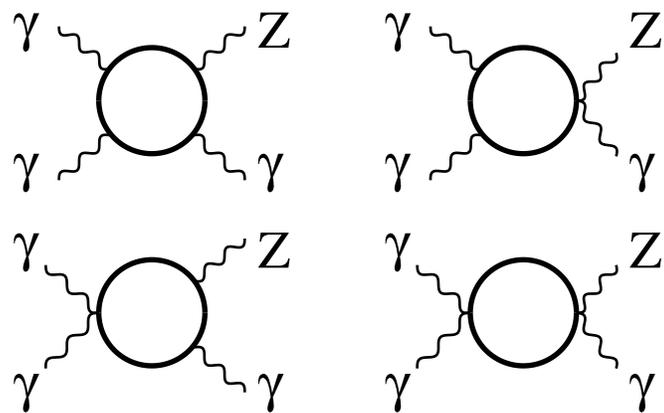
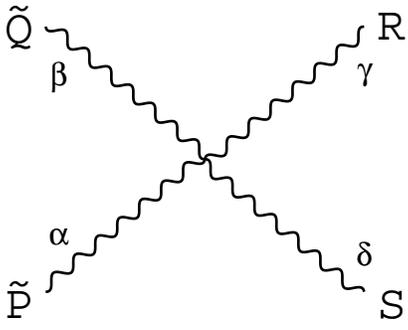
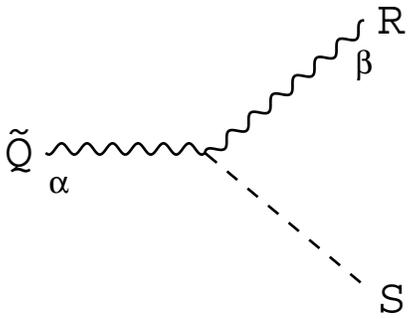
\bye